\documentclass[review]{elsarticle}
\usepackage{amsmath,bm}
\usepackage{graphicx}
\usepackage{amsmath}
\usepackage{amsthm}
\usepackage{float}
\usepackage{epsfig,graphics,subfigure,psfrag,amsmath,amssymb}
\usepackage{pdfpages}
\usepackage{subfigure}
\usepackage{graphicx}
\usepackage{booktabs}
\usepackage{moreverb}

\usepackage{hyperref}
\usepackage[hyphenbreaks]{breakurl}
\usepackage{url}

\usepackage{longtable}
\usepackage{array}

\usepackage{multirow} 
\usepackage{amsmath}
\usepackage{xcolor}
\usepackage{algorithmic}
\usepackage[vlined,boxed,ruled]{algorithm2e}

\newcounter{thm}
 \newcounter{ex}
 \newcounter{re}
\journal{Journal of \LaTeX\ Templates}

\begin{document}

\begin{frontmatter}

\title{Research on the Security of Blockchain Data: A Survey}



\author[a]{Liehuang Zhu}

\author[a,b]{Baokun Zheng}

\author[a]{Meng Shen\corref{mycorrespondingauthor}}
\cortext[mycorrespondingauthor]{Corresponding author}

\ead{shenmeng@bit.edu.cn}

\author[a]{Feng Gao}

\author[a]{Hongyu Li}

\author[a]{Kexin Shi}

\address[a]{School of Computer Science \& Technology, Beijing Institute of Technology, Beijing, China.}
\address[b]{School of Information Management for Law, China University of Political Science and Law, Beijing, China.}

\begin{abstract}
With the more and more extensive application of blockchain, blockchain security has been widely concerned by the society and deeply studied by scholars. Moreover, the security of blockchain data directly affects the security of various applications of blockchain. In this survey, we perform a comprehensive classification and summary of the security of blockchain data. First, we present classification of blockchain data attacks. Subsequently, we present the attacks and defenses of blockchain data in terms of privacy, availability, integrity and controllability. Data privacy attacks present data leakage or data obtained by attackers through analysis. Data availability attacks present abnormal or incorrect access to blockchain data. Data integrity attacks present blockchain data being tampered. Data controllability attacks present blockchain data accidentally manipulated by smart contract vulnerability. Finally, we present several important open research directions to identify follow-up studies in this area.

\end{abstract}


\end{frontmatter}


\section{Introduction}\label{1}

Blockchain \cite{NS08} technology combines multiple computer technologies such as encryption, distributed storage, consensus, Peer to Peer (P2P) network, and smart contracts \cite{AA14}. These key technologies make blockchain open, secure, trust and smart. Moreover,  these techniques allow transactions to be continuously linked to blockchain. Blockchain records all transactions and historical data by establishing a jointly maintained and untampered database.  Internet users who do not know each other can reach a credit agreement through smart contracts, point-to-point ledger, or digital encryption without any central trust \cite{PM15}. Thereby blockchain has attracted extensive research attention from various industries \cite{P17,BW17,G17,BGJ+17,ZLZ+15,W16}.

With the development of blockchain applications, the security of blockchain data is particularly important, and it is the fundamental enabling factors of many blockchain applications. Currently, attackers use the characteristics of the blockchain itself to conduct various attacks on the blockchain data, which makes the blockchain data face various threats. (1) The openness of blockchain data exposes users' privacy. Attackers can find the relationship between addresses through transactions \cite{RH12}. (2) Network attacks cause abnormal or incorrect access to blockchain data, which undermine the availability of the blockchain data. A Bitcoin address can be associated with an Internet protocol (IP) address; therefore, Attackers can track the correspondence among addresses, users and real identity \cite{KPD+14,BAD+14}. (3) Blockchain data will be tampered if an attacker passes an attack on the blockchain consensus mechanism. Blockchain is also vulnerable to selfish mining attacks \cite{BL13,CNL14,ES14}. These undermine the integrity of blockchain data. (4) Smart contract vulnerabilities can cause serious problems by making blockchain data not controlled by users \cite{LDH16}. In addition to these enumerated problems, there are many other security threats in blockchain, such as mining pool attacks \cite{BL13} and miner attacks \cite{R11}. These threats seriously affect the data security of blockchain, which threatens the related blockchain applications.

There are some recent surveys about blockchain security. Gervais et al. \cite{AGV+16} surveyed security and adversarial strategies of proof of work (PoW). Atzei et al. \cite{ABC17} and Luu et al. \cite{LDH16} investigated the vulnerabilities of smart contracts. However, these surveys focused on the security and privacy of a certain aspect of blockchain. Li et al. \cite{LJC17} summarized some cases of attacks against blockchain 1.0 and 2.0. However, their article  lacks systematic description and categorization of threats and countermeasures. With the rapid development of blockchain, many new threats and countermeasures have emerged, and up-to-date research is needed to meet the needs of blockchain development.

Our paper summarizes and analyzes the security of blockchain data. We present the attacks and defenses of blockchain data in terms of the privacy, availability, integrity and controllability. Data privacy attacks include the threats of transaction and identity privacy. Data availability attacks include the threats brought by network traceability and eclipse attacks. Data integrity attacks include the threats brought by double-spending, selfish mining, and block withholding attacks. Data controllability attacks include the vulnerabilities of smart contracts. We present corresponding countermeasures for the threats for each type of attack.

Our main contributions include the following:

\begin{itemize}

\item We created a comprehensive classification and summary of the security of blockchain data.
\item We present the attacks and defenses of blockchain data in terms of the privacy, availability, integrity and controllability.
\item We describe attacks and defenses in a contrastive way. According to blockchain data security features, we present the attack mechanism and the countermeasures. We present the evolution of the attacks and countermeasures.
\item We discuss research hotspots and present several future research directions.

\end{itemize}

The rest of this paper includes: In section \ref{2}, we introduces classification of attacks. In section \ref{3}, we present blockchain data challenges. In sections \ref{4}, we present corresponding solutions. Finally, we present research directions for the future in section \ref{5} and summarize our survey in section \ref{6}.

\section{Classification of Blockchain Attacks}\label{2}

\begin{table*}[!t]\scriptsize
\renewcommand{\arraystretch}{1.3}
\caption{Classification of Blockchain Attacks}
\label{tab1}
\centering
\begin{tabular}{m{1.4in} m{1.3in} m{0.8in} m{0.6in}}
\hline
Data privacy attacks & & & \\
\hline
Transaction privacy attacks & Identity privacy attacks & & \\
\hline
\cite{RH12,LZD+16,RS13} & \cite{RH12,LZD+16,RS13,AKR+13,M15,MPJ+13,Z14} & & \\
\hline
Data availability attacks & & & \\
\hline
Network traceability attacks & Eclipse attacks & & \\
\hline
\cite{GAN17,BJA+15,ZAB15,KPD+14,BAD+14} & \cite{EAA+15,SAT+06,BAD+14,K14,R14,BFB14,CVE14} & & \\
\hline
Data integrity attacks & & & \\
\hline
Double-spending attacks & Selfish mining attacks & Block withholding attacks &  \\
\hline
\cite{AJS+00,EST10,BC18,R14,GOE12,PR16} & \cite{ES14,NL14,SSZ16,NKM+16,CKW+16} & \cite{R11,ES14,NL14,E15,YDY17} & \\
\hline
Data controllability attacks & & & \\
\hline
Logic problems & Semantic misunderstandings & Design problems verifier's dilemma & Privacy preservation \\
\hline
\cite{KMA16} & \cite{LDH16} & \cite{LJR15} & \cite{KMS16,ZCC16} \\
\hline
\end{tabular}
\end{table*}

As previously mentioned, the security of blockchain data is mainly divided into four aspects: privacy, availability, integrity, and controllability. Each aspect contains several attacks, as shown in Table \ref{tab1}. Data privacy attacks refer to data leakage or data obtained by attackers through analysis. In this aspect, we present the threats of transaction and identity privacy. Data availability attacks refer to abnormal or incorrect access to blockchain data. In this aspect, we present the threats brought by network traceability and eclipse attacks. Data integrity attacks refer to blockchain data being tampered. In this aspect, we present the threats brought by double-spending, mining pool, and miner attacks. Data controllability attacks refer to blockchain data accidentally manipulated by smart contract vulnerability. In this aspect, we present the vulnerabilities of smart contracts.

\section{Blockchain Data Attacks}\label{3}

\subsection{Data Privacy Attacks}

Using the transaction process of Bitcoin as an example, we can analyze the threats and corresponding solutions of blockchain data. In Bitcoin, every transaction is traceable. A transaction output is the input to another transaction, thus forming a transaction chain. Based on the chain of transactions, the analyst can obtain the use of any coins and the relevant transactions of any address.

\begin{figure}[t]
  \centering
  \includegraphics[width=3in]{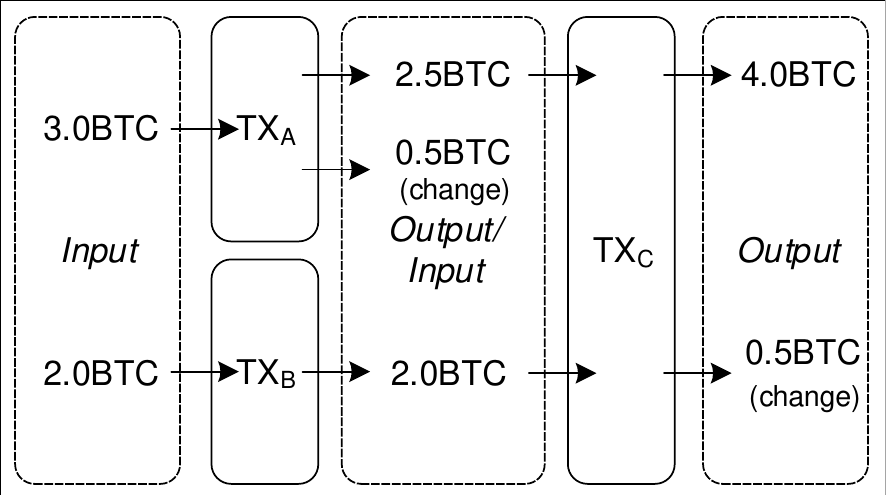}
  \caption{Bitcoin transactions schematic diagram.}
  \label{fig1}
\end{figure}

As shown in Figure \ref{fig1}, an example illustrates the simple transaction process, assuming that Alice launched transaction $A$, Bob launched transaction $B$, and Mike launched transaction $C$. The relationship of transaction input and output is as shown in the figure.

\begin{figure}[t]
  \centering
  \includegraphics[width=3.5in]{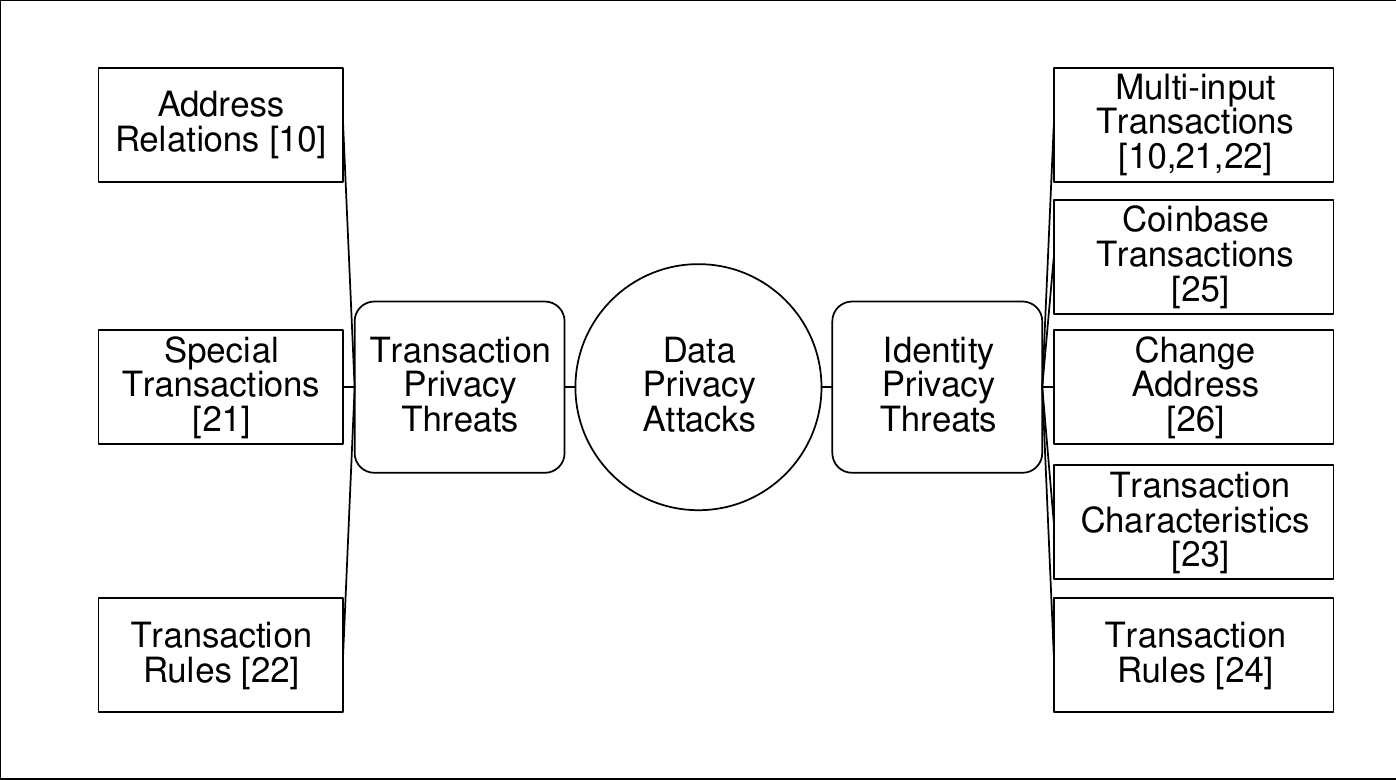}
  \caption{Data privacy attacks.}
  \label{fig2}
\end{figure}

Potential attackers can analyze users' transaction and identity privacy by analyzing transaction records, as shown in Figure \ref{fig2}.

\subsubsection{\textit{Transaction Privacy Attacks}}

The transaction input comes from the output of another transaction. Based on the blockchain of transactions, analysts can obtain the following information:

\begin{itemize}

\item \textit{The Use of Bitcoin:} The Bitcoin came from the mining process, which was first recorded in the miners' mining address and then transferred to other addresses. Both mining and transaction information will be recorded in the global ledger. Therefore, by analyzing these public data, an attacker can acquire all transactions of any Bitcoin.
\item \textit{The Blockchain Transaction Addresses:} Each blockchain transaction details the information of all the input and output addresses.

\end{itemize}

Therefore, analysts can obtain the following information:

\begin{itemize}

\item \textit{Finding Bitcoin Relations between Different Addresses:} The transfer of the coins between accounts reflects the relationship between accounts. Reid and Harrigan \cite{RH12} analyzed the accounts published by WikiLeaks and tallied the balances , Bitcoin sources and flows of the Bitcoin addresses published on the WikiLeaks website. The paper also analyzed a stolen address in Bitcoin and found the five closest addresses to the theft address, revealing the pre- and post-theft Bitcoin flows.
\item \textit{Tracking Special Transactions:} We can monitor the transaction information of special transactions involving large or suspected malicious acts such as theft, and further trace the flow of Bitcoins through continuous observation. Liao et al. \cite{LZD+16} showed the attack of CryptoLocker that extorts Bitcoin by encrypting the victim's files. The authors studied the relation transactions of public Bitcoin ransomware addresses. 968 addresses that belong to the organization were found, and ransom transactions worth 1128.40 BTC were identified. This information assists in determining the identity of the criminals.
\item \textit{Discovering the Rule of Transactions:} The Rule of Transactions can reveal the relationship between transactions. Ron and Shamir \cite{RS13} focused on the transactions statistics. They traced 364 transactions of more than 50,000 BTC and studied the transactions rules for a transaction of 90,000 BTC . The authors found that the large transactions used a variety of methods to disperse Bitcoins to different addresses. These transaction modes include long chains, fork-merge patterns and self-loops, keeping Bitcoins in $saving\;accounts$, binary tree-like distributions.

\end{itemize}

\subsubsection{\textit{Identity Privacy Attacks}}

There are many clues and side information in blockchain transactions, and it is possible to use these clues and side information to speculate about the identity privacy.

The architecture of Bitcoin can be revealed the following:

\begin{itemize}

\item \textit{Multiple Input Addresses} belong to the same person or organization. The multiple input transactions are initiated by the same user because each input in a multi-input transaction requires a separate signature \cite{RH12,LZD+16,RS13}.
\item \textit{Multiple Output Addresses} in the same Coinbase \cite{BC18} transaction belong to the same user set. Many miners want to increase their income by joining one mining pool where they participate in collective mining. All miner addresses involved in mining are recorded as the Coinbase transaction output.
\item \textit{Input Addresses and Change Address} belong to the same user. The change address is generated by the Bitcoin system, which save the change Bitcoins in one transaction \cite{CHANGE18}. The features of the change address include the following: the status of the output address is usually only once, the change address only belongs to the transaction input or output in one transaction; and only change address cannot appear in the output address.

\end{itemize}

The first clues and side information are due to the design of Bitcoin itself. By using these clues and side information, analysts can discover the correlation between different addresses and reduce the anonymity of blockchain addresses. Meiklejohn et al. \cite{MPJ+13} used heuristic analysis to analyze transaction data in the blockchain to identify the same user's different addresses. They analyzed the public addresses of Silk Road and the addresses associated with some theft cases and found many related addresses. Zhao et al. \cite{Z14} proposed a clustering process for Bitcoin transaction data. Based on the analysis of 35,587,286 addresses in the global ledger of Bitcoin, there were 13,062,822 different users.

The second clues and side information is the following:

\begin{itemize}

\item \textit{Transaction Characteristics:} The transaction characteristics are usually related to the actual transaction processes. Many transaction behaviors have their own characteristics in daily life \cite{AKR+13}. For example, transactions at breakfast stores often occur in the morning, and the transaction amount is set at one to 20 coins. The gas station transaction time is an average, but the transaction amount is concentrated in a few specific values, i.e., 100 coins, 200 coins, or full price (changes based on the change of oil price have universal regularity).
\item \textit{Transaction Rules:} Each user has a different transaction behavior. Monaco \cite{M15} analyzed the transaction parameters, and then proposed a method based on parameter identification.

\end{itemize}

\subsection{Data Availability Attacks}

The main threat of data availability is to make abnormal or incorrect access to blockchain data.

\subsubsection{\textit{Network Traceability Attacks}}

In Bitcoin network, IP address, topology, and transmission information can be obtained by attackers. Based on this information, the analyst can analyze user identity privacy. Each network node is connected to many other nodes through the P2P network, and the connection relationship between these nodes can be analyzed. \cite{BAI15,AJA+15}.

\begin{figure}[t]
  \centering
  \includegraphics[width=3.5in]{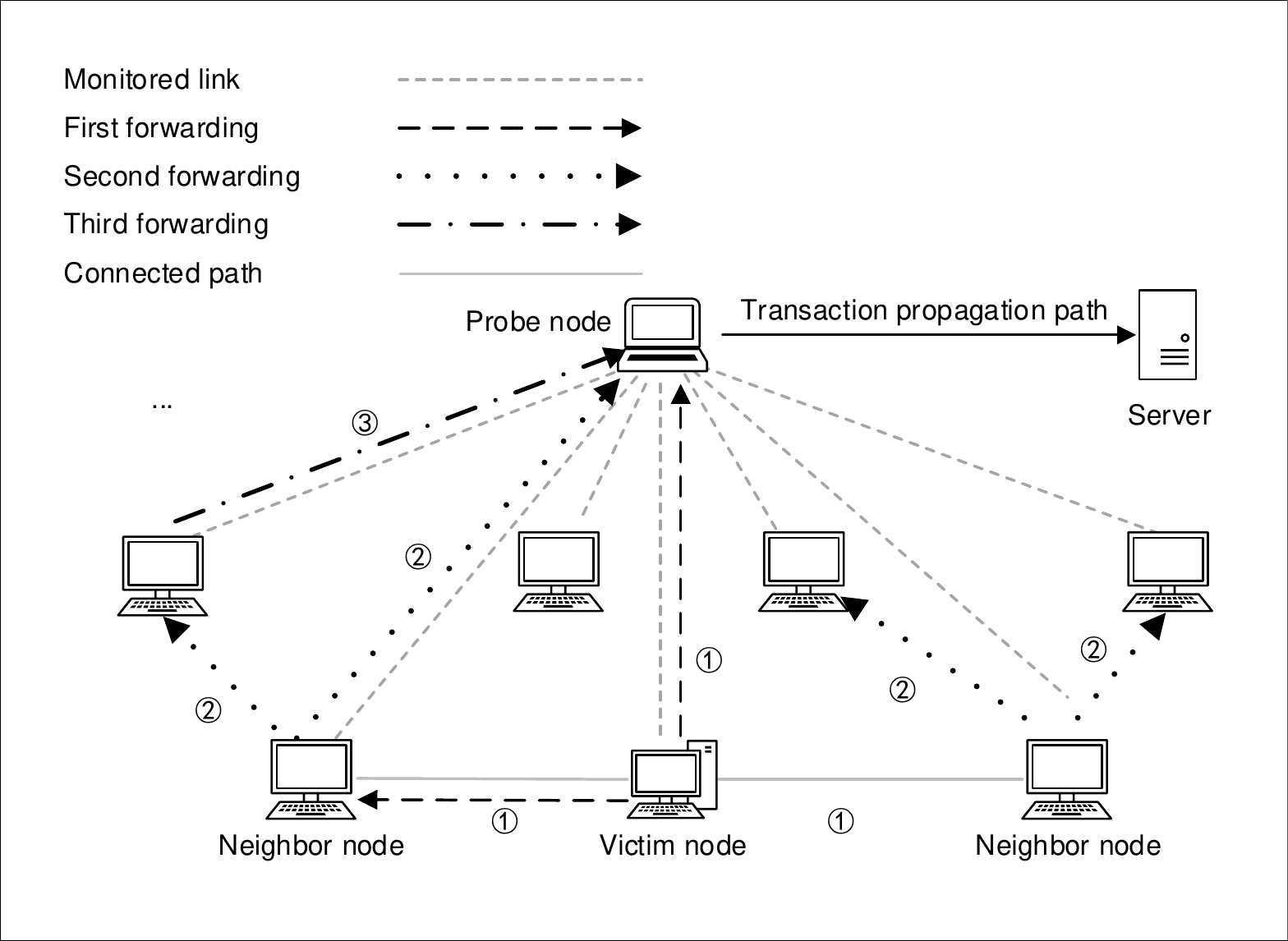}
  \caption{Transaction traceability mechanism diagram.}
  \label{fig3}
\end{figure}

Transaction traceability is to estimate the transaction propagation path according to the time order that the different nodes send the transaction to arrive at the probe, as shown in Figure \ref{fig3}. Ideally, the originating node is the earliest to arrive at the probe, the neighbor arrives at the neighbor second, and the order the next $n-th$ neighboring nodes arrive at the probe will increase with distance. In the actual environment, the time order of different nodes' transmitted transactions arriving at probes is affected by many factors, such as network delay and delay transmitting strategy, and the transactions transmitted by the long-distance nodes may arrive earlier. To accurately analyze the matching degree of transaction ranking and node network topology, we will consider a variety of influencing factors and calculate the trading order accuracy.

Network traceability technology uses the collection of Bitcoin network transmissions of information to analyze the transmission path of Bitcoin transactions in the network, then tracks the transaction-generated server IP information. This technology can directly contact an anonymous transaction via the trade originating node's IP address to permit traceability. However, the existing network traceability technology has a low accuracy and generally needs additional computing and storage resources; therefore, it is less practical.

Bitcoin users can make double-sided transactions of Bitcoin tokens by creating Bitcoin transactions with other users on servers anywhere in the world \cite{GAN17,BJA+15,ZAB15}. Since the transaction does not require the participation of third parties, and the addresses used by both parties in the transaction are anonymous, the real identity of the Bitcoin traders is hard to find.

Transaction traceability technology desires to track the transmission path of Bitcoin transactions in the network to find the originating node of a transaction, which is the first server node of the transaction in the Bitcoin network. Once a Bitcoin transaction is associated with the IP address of the originating node, the anonymous account number in the transaction can be associated with the user identity, which helps to identify the identity information of the malicious trader and analyze the flow of the Bitcoin funds.

Network traceability technology is to analyze the transaction information transmitted by the Bitcoin network, locate the propagation path of a specific transaction, and then infer the origin node of the transaction.

Koshy et al. \cite{KPD+14} analyzed the patterns of Bitcoin transactions in the network and found that we can search for the origin node by using the special transaction mode. For example, a transaction that is transmitted only by one node is usually due to a problem with the transaction format, then this transaction is transmitted only once by the originating node. However, the effect of this method is limited due to the small proportion of all transactions in the special transaction mode (less than 9\% of the special transactions in the paper trial).

Biryukov et al. \cite{BAD+14} analyzed transaction traceability using neighbor nodes. By using neighbor nodes as the basis for judgment, the accuracy of traceability can be improved. However, the solution must continuously send information to all the nodes in the Bitcoin network, which may cause serious interference to the Bitcoin network and is less practical.

\subsubsection{\textit{Eclipse Attacks}}

Heilman et al. \cite{EAA+15} described eclipse attack, which exploits the broadcast features of P2P networks to attack. The attacker controls the reception and transmission of all information of the victim node, causing the victim node's inbound connection to the illegal node.

The attack node maliciously fills the victim node's routing table before the victim node of the blockchain restarts, forcing the victim node to restart and establish an outgoing connection with the attack address in the routing table \cite{EAA+15} \cite{SAT+06}. At the same time, the attack node continuously establishes an incoming connection with the victim node. Eventually, the channel of the monopolistic victim node is reached, and the purpose of its information flow is controlled so that it can only receive useless or even malicious information sent by the attack node. If the attack node can successfully implement eclipse attacks on more nodes, it can control the blockchain channel and information flow of more nodes, and gradually control most of the blockchain networks. Attackers can even launch 51\% attacks and double-spending attacks on this basis, causing more serious consequences.

The eclipse attack process is usually divided into four steps, as shown in Figure \ref{fig4} and detailed as follows:

\begin{figure}[t]
  \centering
  \includegraphics[width=3.5in]{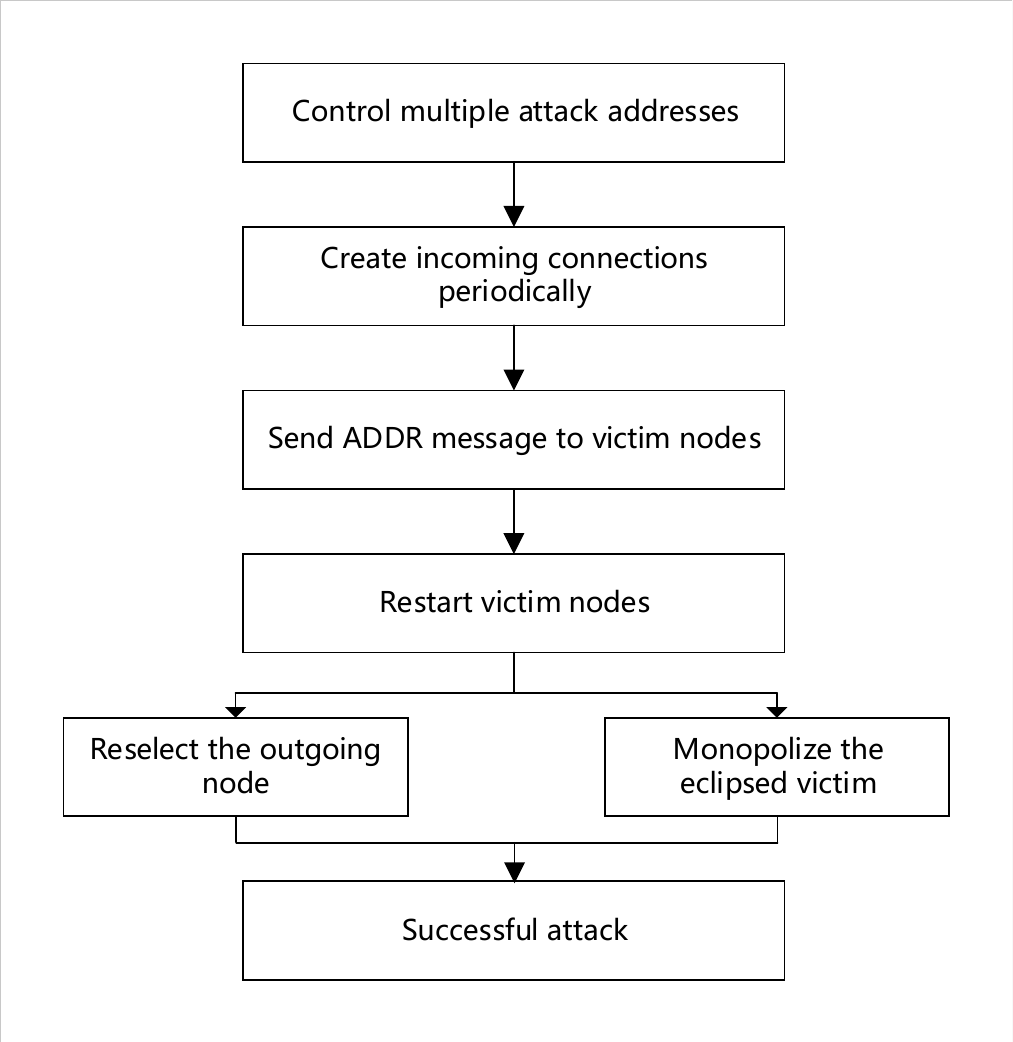}
  \caption{Eclipse attack flow diagram.}
  \label{fig4}
\end{figure}

\begin{itemize}

\item \textit{Populating Tried and New Tables:} The blockchain node is capable of receiving the addresses of the unsolicited incoming connections and the unsolicited ADDR message. The addresses of the incoming connections will be stored in the tried table, and the addresses contained in the ADDR message can be inserted directly into the new table. The nodes will not test the addresses' connectivity. Therefore, when the attack nodes are connected to the victim node through the attack addresses, the attack node can send an ADDR message containing a lot of ``trash" IP addresses that will gradually overwrite all legal addresses of the new table. The nodes rarely get network information from their neighbor nodes and DNS seeders. Therefore, when an attacker overwrites the tried and new tables of the victim node, the victim node almost never verifies its authenticity by querying a legitimate peer or sower.

\item \textit{Restarting the Victim:} The victim node will be restarted by the eclipse attack. After the node is restarted, the victim node can be connected to the attack addresses. The reasons for the Bitcoin node restart include: ISP shutdown, machine shutdown, operating system upgrade of the mining machine, etc \cite{BAD+14,K14,R14,BFB14,CVE14}.

\item \textit{Selecting Outgoing Connections:} After the victim node is restarted, if the address is selected from the new table to establish an outgoing connection, all connections fail. Therefore, the victim node is forced to only pick the addresses from the tried table. Because the victim node prefers to choose the updated addresses, all the outgoing connections of the victim node are connected to the attack addresses.

\item \textit{Monopolizing the Eclipsed Victim:} If the above attack is successful, The attacker must control all incoming connections to the victim node in order to truly monopolize the victim node.

\end{itemize}

In addition, the eclipse attack will also cause other attacks. The main attacks are shown in Table \ref{tab4}, the details of the attacks will be described in detail in section \ref{5}.

\begin{table*}[!t]\footnotesize
\renewcommand{\arraystretch}{1.3}
\caption{Some other Attacks that may be caused by an Eclipse Attack}
\label{tab4}
\centering
\begin{tabular}{|m{1.1in}<{\centering}|m{1.1in}|m{2in}|}
\hline
Reference & Attacks & Description  \\
\hline
\cite{AJS+00,EST10,BC18,GOE12,R14,PR16}  & Double-spending threat & Using the same cryptocurrency in multiple transactions by a sender  \\
\hline
\cite{ES14,NL14,SSZ16,NKM+16,CKW+16} & Selfish mining attack & Hiding the excavated blocks to cause the chain to fork \\
\hline
\cite{ES14,NL14,R11,E15,YDY17} & Block withholding attack & The attacker never submits any blocks \\
\hline
\end{tabular}
\end{table*}

\subsection{Data Integrity Attacks}

Data integrity attacks primarily include double-spending, selfish mining, and block withholding attacks.



\subsubsection{\textit{Double-Spending Attacks}}

Double-spending attacks refer to the use of the same cryptocurrency in multiple transactions by a sender. Bitcoins use the PoW system, which has approximately ten minutes of confirmation time between blocks. Therefore, an attacker will implement a the attack in this time interval. In addition, if the attacker has a significant amount of computational power, he or she will be more likely to successfully perform the attack.

Double-spending attack is a unique attack on the bitcoin system that falls into two types:

\begin{itemize}

\item \textit{Attacker Used the same Bitcoin to Trade with multiple Users at the same Time:} If these trading users complete the transaction without the transaction being recorded in the legal blockchain, the attacker achieves the goal of double spending or even multiple spending \cite{R14,AJS+00,EST10,BC18+,GOE12}. Although in the multiple transactions launched by the attacker, only one transaction is considered legal and recorded in the blockchain, the transaction has been completed and the attacker has benefited from the attack.
\item \textit{Attacker Used his own Computing Power to Launch an Attack:} The attacker used the same Bitcoin to trade both transaction $A$ and transaction $B$ with two users. If the transactions $A$ is confirmed to be recorded in the blockchain, transaction $A$ is completed. Because the attacker has a powerful computing power, he records transaction $B$ in the private blockchain and mines a longer chain than the legal one, prompting transaction $B$ to be confirmed and completing transaction $B$ \cite{PR16}.

\end{itemize}

Ghassan et al. \cite{GOE12} analyzed the double-spending threat of Bitcoin in the fast payment scenario. Figure \ref{fig5} illustrates the attack model. We assume that the attacker $A$ must pay BTCs to a vendor $V$, and $A$ creates the transaction $TX_v$ to $V$. Simultaneously, to realize double-spending, $A$ creates another transaction $TX_a$ that has the same BTCs as those involved in $TX_v$'s inputs. The successful implementation of a double-spending attack must meet the following three requirements:

\begin{figure}[t]
  \centering
  \includegraphics[width=3in]{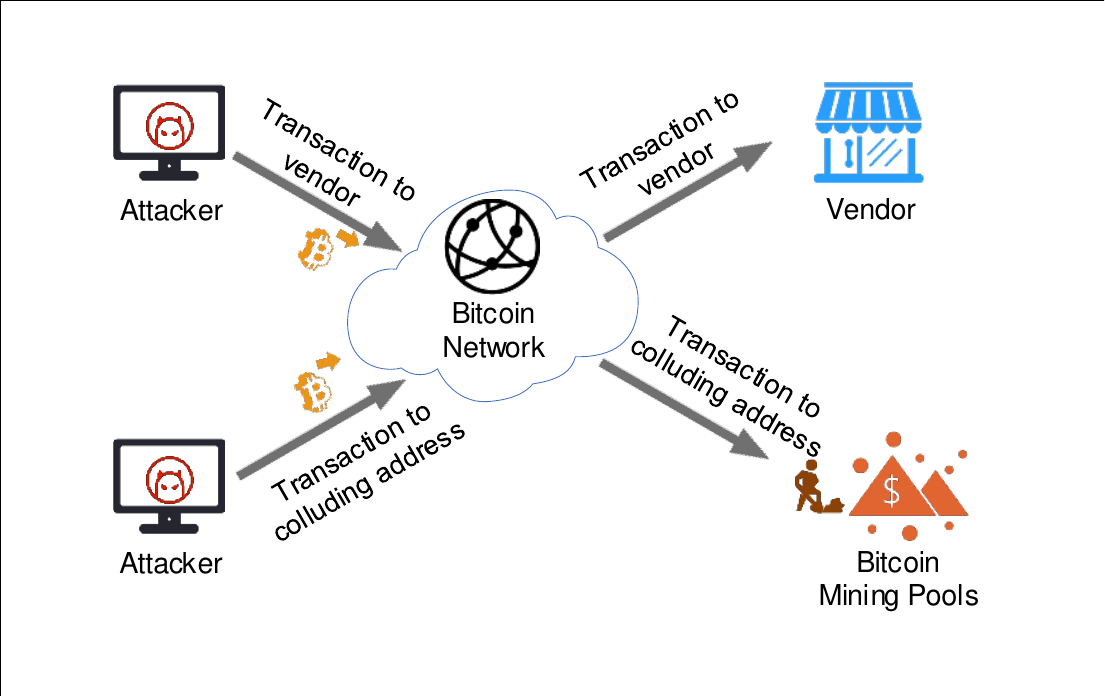}
  \caption{Double-spending attack model.}
  \label{fig5}
\end{figure}

\begin{itemize}

\item $TR_v$ is added to $V$'s wallet.
\item $TR_a$ is confirmed in the blockchain.
\item $V$'s service time is less than the time when $V$ detects the wrong behavior.

\end{itemize}

In fact, because of the PoW mechanism, it usually takes ten minutes to prevent double-spending attacks, so it does not apply to quick payment scenarios. Additionally, without a suitable detection mechanism, double-spending attacks can be implemented in a low-cost manner.

\subsubsection{\textit{Selfish Mining Attacks}}

Selfish mining attack \cite{ES14} is a typical attack on blockchain. A cryptocurrency like Bitcoin requires a high computing power to solve the cryptographic problem for a miner, so the mining becomes very difficult. In view of this, a group of miners (or mining pools) are usually combined with each other and share the rewards after successfully solving the password problem. This helps individual miners produce more continuous and constant income when mining alone.

Eyal et al. \cite{ES14} proposed that if there is a group of selfish miners who use selfish mining strategies and succeed, it may invalidate the work of honest miners. A malicious mining pool does not publish the blocks it finds and creates a fork. Therefore, there are public chain maintained by honest miners and the private fork by malicious mining pools. Since the fork is the longest chain in the current network, it will be recognized as a legal chain by honest miners. Therefore, the original public chain and the honest data it contains will be discarded. The results of the study show that usually the selfish mining strategy will get more benefits. At the same time, the analysis shows that if the selfish pool exceeds one third of the total net, the existing protocol will no longer be safe.

Courtois et al. \cite{NL14} conducted experimental simulation and theoretical analysis of selfish mining. The results show that the computational waste of Bitcoin is minimal, and it is even decreasing over time. Sapirshtein et al. \cite{SSZ16} studied the optimal strategy of the selfish-mining underlying model. Nayak et al. \cite{NKM+16} shows when this attack is combined with an eclipse attack, these strategies sometimes result in a gain of 30\% depending on the different parameters. Carlsten et al. \cite{CKW+16} proposed a more complex selfish mining strategy that led to uneven returns and exceeded default mining and traditional selfish mining. Once deployed, the attack will be profitable, which could result in 51\% of attacks or consensus failures.

\subsubsection{\textit{Block withholding Attack}}

\begin{figure}[t]
  \centering
  \includegraphics[width=3in]{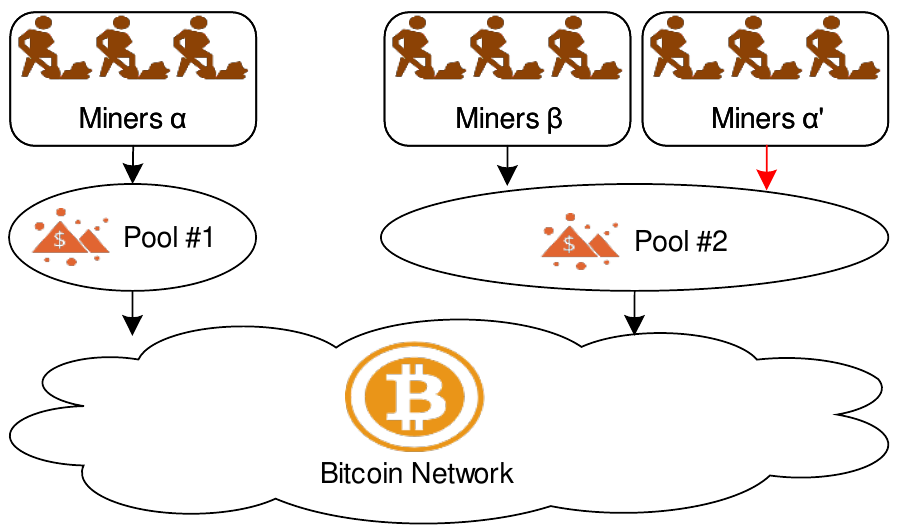}
  \caption{Block withholding attack diagram.}
  \label{fig6}
\end{figure}

Block withholding attack \cite{R11} is one of the typical attacks on blockchain. In the attack, some malicious attackers who have joined the joint mining pool do not have any mining blocks, which reduces the revenue of the mining pool and wastes the computing power provided by other miners. This kind of attack is also called sabotage attack. Usually, the malicious miners will not have any benefit. The blocker attack will cause different losses to the miners and mining pools, and the mining pools' losses are relatively large compared to the miners' low cost. As a result, block withholding attack is more common in competing mining pools and less common in miners  \cite{ES14}. Block withholding attack diagram is shown in Figure \ref{fig6}.

Courtois et al. \cite{NL14} analyzed the actual examples and found that the main hazard of the malicious miners who can profit from this attack is to waste the computing resources of the mining pools and reduce the income of the mining pools. Eyal \cite{E15} analyzed the game of miners' dilemmas, and there is a balance between competing mining pools, which makes miners repeatedly choose whether to attack. Kwon et al. \cite{YDY17} extended the BWH attack \cite{E15} and proposed a new attack method called fork after withholding (FAW) attack. The attack uses selfish mining attacks based on BWH attacks. The FAW attacks'  frequency is four times as often as BWH attacks, and the attack will also gain an additional 56\%. In addition, the research showed that when two pools attack each other, the more computing power, the easier it is to win.

\subsection{Data Controllability Attacks}

Smart contracts are blockchain-based programs that directly control digital assets. Additionally, it is executed automatically by a computer system. Nick Saab \cite{KMA16} indicates that smart contracts are essentially a recognized tool for forming relationships between individuals, institutions and property, i.e., a set of agreements that form relationships and reach consensus.  Smart contracts work similar to if-then statements. When a predefined trigger condition is reached, the smart contract is executed.

The most significant feature of smart contracts based on Ethereum is Turing completeness. Smart contracts are written into the blockchain in the form of digitization, which is protected by the blockchain. The entire process of storing, reading and executing by the characteristics of blockchain is transparent and traceable and cannot be changed. If a user wants to modify a smart contract, he or she must control at least 51\% of the calculation power.

\begin{figure}[t]
  \centering
  \includegraphics[width=3.5in]{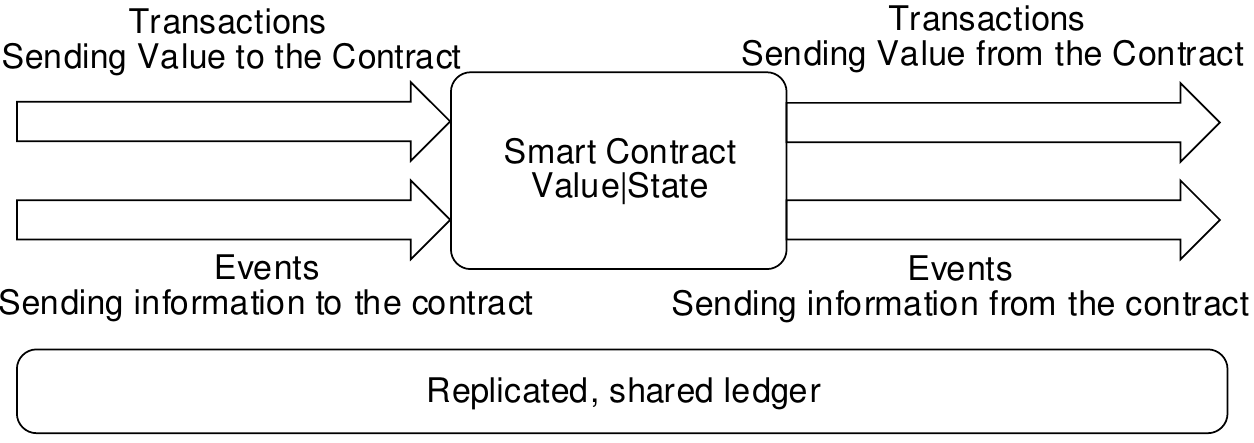}
  \caption{Smart contract schematic diagram.}
  \label{fig7}
\end{figure}

The input to a smart contract includes transactions and events. The transactions mainly include transaction data, and the events refer to the description of the transactions. Smart contracts are triggered when conditions are met. Smart contracts exist to allow a complex set of digitized promises with triggering conditions to be properly executed according to the wishes of the participants, as shown in Figure \ref{fig7}.

\begin{table*}[!htb]\footnotesize
\caption{Smart Contract Vulnerability Summary}
\label{tab5}
\footnotesize
\centering
\begin{tabular}{|m{0.6in}<{\centering}|m{1.5in}|m{2in}|}
\hline
Reference & Classification & Description \\
\hline
\cite{KMA16} & Errors in encoding state machines & Coins may not be returned. \\
\hline
\cite{KMA16} & Failing to use cryptography & Attackers can adjust their input according to the user's input. \\
\hline
\cite{KMA16} & Misaligned incentives & The second user will no longer have any action (i.e., no reward) when he feels he may have failed. \\
\hline
\cite{KMA16} & Ethereum-specific mistakes  & The problem causes some bugs. \\
\hline
\cite{LDH16} & Transaction-ordering dependence  & If the transaction verification order is different, the execution result is different. \\
\hline
\cite{LDH16} & Timestamp dependence  & Miners can attempt different timestamps in advance to calculate the winning value, to award the prize to the winners they desire. \\
\hline
\cite{LDH16} & Mishandled exceptions  & If there is an error during the call, it will return to the pre-contract state. \\
\hline
\cite{LDH16} & Reentrancy vulnerability  & The new chain is broadcast to the old chain, and the transaction is still successful, resulting in confusion. \\
\hline
\cite{LJR15} & Resource exhaustion attack by problem givers & The miners were unable to mine the next block, causing damage to the miners. \\
\hline
\cite{LJR15} & Incorrect transaction attack by provers & The user not only wastes his or her reward but also does not receive a correct answer. \\
\hline
\cite{KMS16} & Data feed  & An attacker could gain valuable information by analyzing the transaction history. \\
\hline

\end{tabular}
\end{table*}

Smart contracts maximize open source and solve the trust issues of traditional $SAAS$ services. Developers can validate the code's original code through compilation to prove the code's usability. Therefore, smart contracts are applied in various fields such as housing leasing, savings insurance, financial lending, and probation.

\begin{figure}[t]
  \centering
  \includegraphics[width=3.5in]{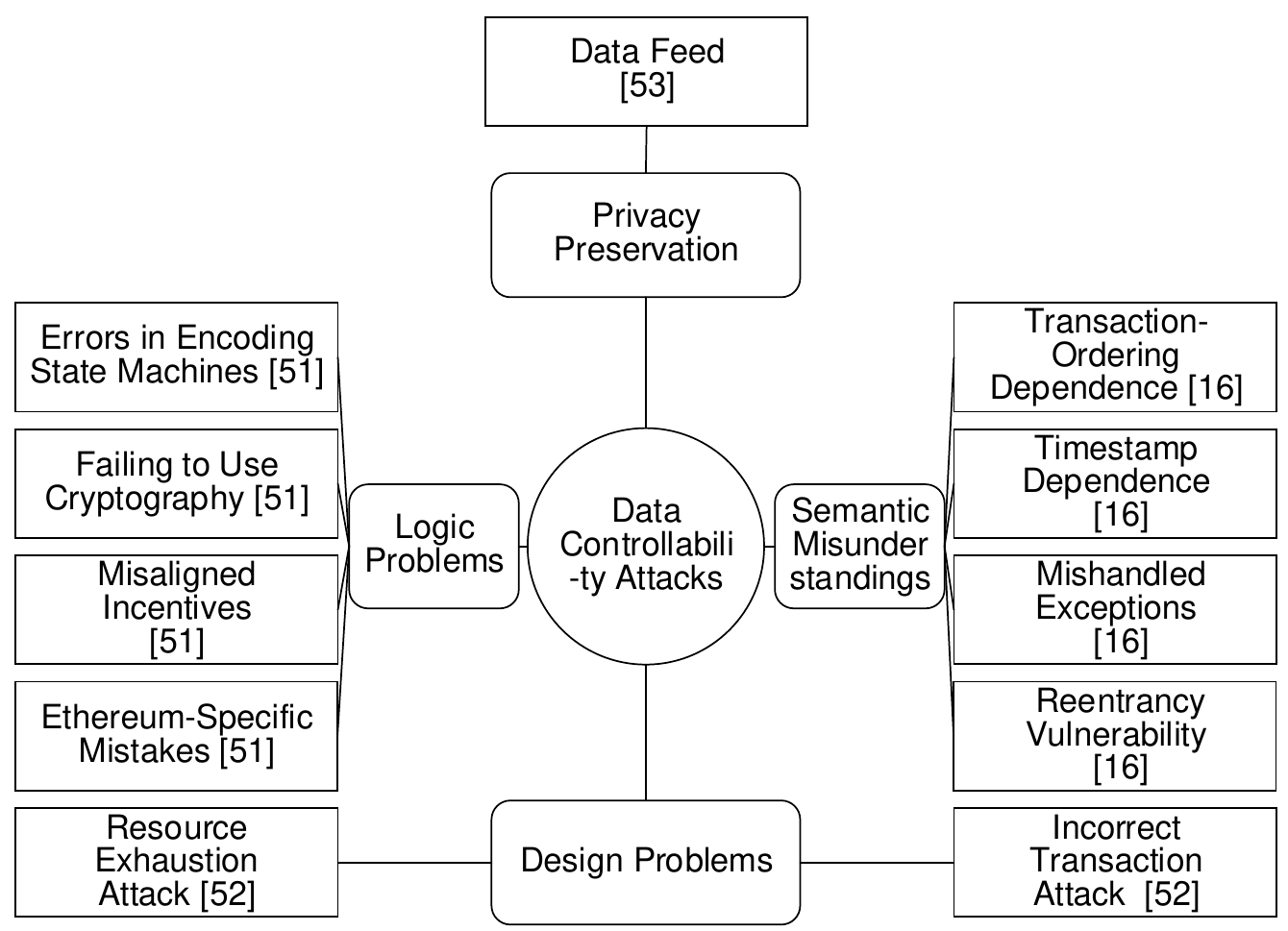}
  \caption{Data controllability attacks.}
  \label{fig8}
\end{figure}

However, smart contracts solve the contradiction between the openness and security of the code, currently, there remain other security privacy issues. A smart contract vulnerability summary is shown in Table \ref{tab5} and Figure \ref{fig8}.

In the aspect of the programming of smart contracts \cite{KMA16}, if the contract itself has a logic trap, the execution result of the smart contract can be changed through certain trigger conditions to generate the result that is beneficial to the developer. Semantic misunderstandings \cite{LDH16} may occur during the execution of a smart contract deployed at Ethereum, inconsistent with the notion of a smart contract written by the original developer, causing the result to deviate from the expected performance. Incentives designed for smart contracts \cite{LJR15} may create a verifier's dilemma where honest miners choose to continue verifying or mining to get rewarded when faced with the more difficult task of verifying transactions.  For the data privacy protection of smart contracts, the privacy protection of data \cite{KMS16} and an authenticated data feed \cite{ZCC16} are elaborated. We will detail each threat and its corresponding security method in a later section. After detailing each of the attacks in Table \ref{tab5}, we develop a solution (except for privacy reasons, because they have no vulnerabilities to attack).

\subsubsection{\textit{Logic Problems of Smart Contracts}}

Delmolino et al. \cite{KMA16} gathered smart contracts written by students, and there are many security issues in the design process. The security of smart contracts is very important in the process of programming them. Even if some malicious miners exist, they will not increase their own profits. They search bugs of contracts, and there will be security pitfalls, although the design and implementation of the contract is very simple. This contract mainly consists of two parts, one for each player to input his or her own results and the other for the contract to determine the winner and send him or her the award.

\begin{itemize}

\item \textit{Errors in Encoding State Machines:} During the game between the two players, the contract may contain some mistakes. If there is a third player to send coins to in the contract, no one's coins will be returned. Even a cautious player cannot avoid this situation. Since developers do not fully consider the status of the execution of each step of the contract or clearly define the clear conversion between the states in the programming of smart contracts, these problems arise.
\item \textit{Plaintext transmission:} Players send their inputs in plaintext in smart contracts is a serious problem. Because the transactions are transmitted through plaintext, malicious players only need to intercept other people's inputs first and adjust their input results according to other players' inputs to maximize their profits.
\item \textit{Misaligned Incentives:} Once a player believes that he or she will lose the game, the player may choose not to continue to send any further messages, thus refusing to pay the winner's rewards and protecting his or her own interests.
\item Ethereum-specific Mistakes

\begin{itemize}

\item \textit{Call-stack Bug:} In EVM, there are three locations for data storage, the stack, temporary memory, and permanent memory. Among them, the stack is the only free data storage area; other areas must pay for gas. However, there are some issues that may arise when using the stack. When the stack depth exceeds a certain limit (Ethereum specifies a stack limit of 1024), a stack overflow will occur, and the caller of the contract will have a loss of property.
\item \textit{Blockhash Bug:} In Ethereum, block.prevhash instruction is used only to calculate the hash value of the 256 recent blocks.
\item \textit{Incentive Bugs:} There are some incentive bugs in Ethereum, for example, selecting the winner of the election by using a hash of the block as a random beacon. However, miners can selectively withhold blocks to deviate from this value to gain an unfair advantage. If miners mine a new block, they can check whether they win. If they do not win, the miners will keep this block until next block is generated; then, they will receive a second chance to see if they win the game.
\item \textit{Other Bugs:} One of the important issues that must be considered when designing a smart contract is scalability, as contractual upgrades are inevitable. In EVM, code is completely unmodifiable and it is impossible to load and execute code in memory. The code and data are completely separated. Currently, the upgrade can only be achieved by deploying a new contract, which may require copying all the code in the original contract and redirecting the old contract to the new contract address. Patching the contract or partially upgrading the contract code in EVM is completely impossible.

\end{itemize}

\end{itemize}

\subsubsection{\textit{Semantic Misunderstandings of Smart Contracts}}

Luu et al. \cite{LDH16} documented several several vulnerability in Ethernet smart contracts. These defects refer to differences between the author's design and actual semantics. We introduce the following four types of security issues and countermeasures:

\begin{itemize}

\item \textit{Transaction-Ordering Dependence:} The vulnerability is that the order of transaction validation affects the execution result of intelligent contracts. If the transaction verification order is different, the execution result is different. The attacker provided a smart contract with a prize wager and promised generous rewards when the user determined the correct solution. When the user submits the answer and the transaction has not been verified, the attacker will initiate another transaction, making the transaction's reward value infinitely close to zero. At this point, two unverified transactions appear in the pool. When the miners verify the transactions, the user's transaction that answers the question after verifying the attacker's transaction has the user's rewards significantly reduced; the attacker seldom pays any bonuses to obtain the right answer. In addition, attackers can make miners prioritize their transactions by increasing transaction fees.
\item \textit{Timestamp Dependence:} The Ethereum states that when the timestamp of a new block generated by the miners is greater than that of the last block and the time difference between the two is less than 900 seconds, the new block is considered valid and its timestamp is legal. Timestamp dependency refers to the implementation of smart contracts that depend on the current timestamp; the results of the execution timestamps vary. If there is a lottery contract, the winning value is calculated from the current timestamp and other variables that are known in advance, and the same code as the lucky number will be awarded. At this point, in the process of mining, miners can attempt different timestamps in advance to calculate the winning value, to award the prize to the winner they desire.
\item \textit{Mishandled Exceptions:} Ethereum use the send tool to call a contract or call the contract function directly. If there is an error during the call, it will return to the pre-contract state. King of the Ether Throne (KoET) is an Ethereum contract that will make you a king or queen, might grant you riches, and will immortalize your name. A user can become the new king by paying a certain amount of ether; part of the ether will be used to pay the previous king's remuneration. In the transfer process of the ether, the smart contract did not check the payment results of the transaction, and once the contract call exception occurs, the incumbent kings may also lose their throne and compensation.
\item \textit{Reentrancy Vulnerability:} Traditional reentrancy attacks refer to the situations in which an attacker sends a packet that has been received by the destination host to achieve deception. While a reentrancy attack on the blockchain is a transaction that has been verified on a chain reappearing on another chain for verification, this attack usually occurs when the blockchains are permanently divergent. The famous TheDao hack, because of the inaccuracy of the intelligent contract code, led to a large vulnerability in the transaction funds. Although permanent divergence is used during a later period, there will be reentrancy attacks. The new chain is broadcast to the old chain, and the transaction is still successful, resulting in confusion.

\end{itemize}

\subsubsection{\textit{Design Problems of Smart Contracts}}

Luu et al. \cite{LJR15} presented the verifier's dilemma, which states that verifiers would make a choice between mining or transaction verification. When the transactions are expensive, miners will decide to bypass their verification and mine to ensure their own profits. As a result, miners are vulnerable to resource depletion or incorrect transaction attacks.

\begin{itemize}

\item \textit{Resource Exhaustion Attacks by Problem Givers:} Honest miners verify new transactions in accordance with the protocols. Therefore, attackers can broadcast a very large number of transactions, causing other miners to waste a significant amount of calculation power to verify the correctness of the transaction. To prevent this, Ethereum establishes a gas (consumption) mechanism. Each transaction requires a gas limit and a gas price. Gas limit is the maximum amount of Gas consumption allowed by this transaction. Gas price is a tip. If the user does not have sufficient Ether to pay  the maximum cost of his or her own settings, the transaction is considered invalid, the previously changed state will be restored, and the consumed gas will not be returned to the user. However, the gas mechanism cannot prevent this scenario. The new block is mined by the attacker. Then, the attacker broadcasts their expensive transactions. Even if the attacker pays considerable transaction fees for the deal, because the fees will be given to the miners who mine a new block regardless of the number of transaction fees, the transaction fees will be returned to attacker's account.
\item \textit{Incorrect Transaction Attack by Provers:} When a user asks for a solution to a difficult problem, malicious respondents may provide an incorrect solution, because the transaction verification requires a significant amount of calculation power; therefore, the majority of miners will selectively bypass verification of this transaction, believe it to be correct, and then broadcast it directly to the network, so that the user not only wastes his or her reward but also does not receive the correct answer.

\end{itemize}

Through these two attacks, it is found that if the malicious miners mine a new block, they can attack without considering the reward. When users create a scale of transactions such as block size, Bitcoin can be repaired through pre-defined standard transactions, but there is no such restriction in Ethereum. They presented a safety model that ensures that the miners who did not perform the agreement received minor rewards, and the honest miners who verified the transaction were not affected by the malicious miners.

\subsubsection{\textit{Privacy-Preserving Smart Contracts}}

The global ledger that stores transaction information is public. An attacker could gain valuable information by analyzing the transaction history, including the specific account balance of funds, transaction details, and the flow of specific funds.

\section{Blockchain Data Security Protection}\label{4}

\subsection{Data Privacy Protection}

To ensure the reliability, non-falsification and distributed consistency of the transaction, a special data structure and consensus mechanism are designed. These mechanisms ensure the maintenance of a uniform, high public trust account in distributed untrusted network nodes; however, these mechanisms also lead to privacy risks. The full ledger not only leaks data privacy but also leaks the relation between the traders who are behind the data and the identity privacy \cite{RH12,LZD+16,RS13,AKR+13,M15}. Therefore, the focus of data privacy protection is to hide the data and the information behind it as much as possible.

We classify different protection mechanisms according to the database privacy protection classification methods, as the following:

\subsubsection{\textit{Data Distortion}}

\begin{figure}[t]
  \centering
  \includegraphics[width=3in]{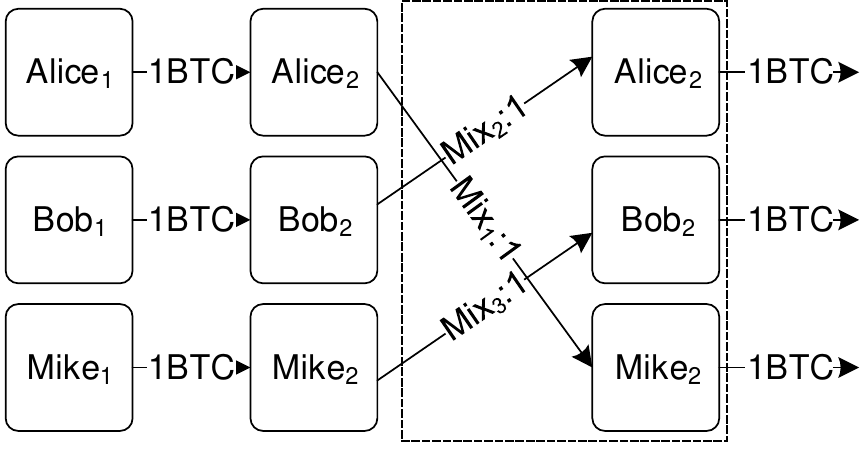}
  \caption{Bitcoin mixing service diagram.}
  \label{fig9}
\end{figure}

Because the blockchain ledger is public, the attacker can find the relationship of the transaction data; the attacker can then infer the transaction and identity privacy. To prevent this attack, we can adopt a method called $mixed\;coin$ \cite{Chaum1981} without changing the transaction results; however, this method adds confusion, as shown in Figure \ref{fig9}. Assuming that Alice, Bob, and Mike have transaction addresses $Alice_1$, $Bob_1$, and $Mike_1$, respectively. In the process of mixing coins, first to generate new addresses $Alice_2$, $Bob_2$, and $Mike_2$ for them, then to send the coins that need to be traded to them, mix the coins by the mixed addresses $Mix_1$, $Mix_2$, and $Mix_3$ and output them to these new addresses $Alice_2$, $Bob_2$, and $Mike_2$, so that others can't grasp the source of these coins.

The $mixed\;coin$ mechanism is classified as follows.

\begin{itemize}

\item \textit{The Mixed Coin Scheme based on a Central Node:}

This scheme utilizes third-party nodes for mixing coins, and the process of mixing coins is done at a third-party node. These methods can improve the security of Bitcoin and other digital currencies without additional technology \cite{bitlaunder,bitcoinfog,blockchain}. However, Its defects include:

\begin{itemize}

\item \textit{Additional Charge and Mixed Coin with a Slower Speed:} A mixed coin service node usually charges a fee. As the mixes increase, the cost rises sharply. The usual delay time is 48 hours, and the transaction costs are between one and three percent.

\item \textit{Risk of Theft:} In this scheme, the third-party node may not perform the agreement after receiving the user's coins and steal the user's coins. The users do not have effective countermeasures.

\item \textit{The Mixing Process is leaked by Intermediate Nodes:} The third-party node in this scheme understands the entire mixed coin process, and the users cannot guarantee that the third-party node will not leak the mixed coin process information.

\end{itemize}

In response to these defects, many improvements have been made. Bonneau et al. \cite{BNM+14} proposed an improved decentralized mixed coin that can be audited. Valenta and Rowan \cite{VR15} designed a blindcoin scheme that can prevent a third party from divulging the process information. Qing et al. \cite{SQY15} presented a blind signature scheme that use the elliptic curve to improve privacy. In 2015, Dash \cite{Dash18}, an anonymous digital coin, launched operation. From an economic point of view, Dash solves the threat posed by centralized mixed coin.

\item \textit{The Mixed Coin Scheme based on Decentralization:}

The program does not depend on third-party nodes. CoinJoin \cite{G13CoinJoin} is the original plan. CoinJoin merge multiple transactions into one transaction in which the relationship of the transaction inputs and outputs can be hidden. For a multi-input and multi-output transaction, a potential attacker cannot effectively distinguish the relation between inputs and outputs by analyzing the transaction information. The idea of CoinJoin is used in many anonymous Bitcoin transactions, for example, Dark Wallet \cite{G14}, CoinShuffle \cite{KyleTorpey18} and JoinMarket \cite{BitcoinTalk18}.

The CoinJoin mechanism enhances the privacy protection capabilities of all users. In a digital currency system, if only a fraction of the nodes uses the CoinJoin agreement, the remainder of the users do not use this protocol, nor do they use the original method.

The CoinJoin mechanism has many defects, as the following:

\begin{itemize}

\item As other users participate, it also faces threats from the other node.

\item The information of each node participating in the mixed coin will be exposed to the other.

\item If some nodes violate the rules, the mixed coin may fail.

\item The parties involved in the mixed coin transaction will be recorded in the ledger.

\end{itemize}

\end{itemize}

Many scholars have proposed solutions. Ruffing et al. \cite{RMK14} propose a completely decentralized CoinShuffle. Based on CoinJoin, the CoinShuffle scheme designs an output address shuffling mechanism. This mechanism can complete the mixing process without a third party, and it can also ensure that the mixed coin participant does not know the relationship. However, the scheme is easy to trigger denial-of-service attacks. Bissias et al. \cite{BOL+14} designed Xim that adopts a multi-wheel and two-square mixed coin agreement. CoinParty \cite{ZGH+15} adopts a secure multiparty computation protocol to implement an improved scheme that can guarantee the effectiveness of the mixed coin process in the case of malicious operation or failure of some hybrid nodes. Monero \cite{Monero18} is a new digital coin with a main characteristic of privacy protection. It adopts the ring signature mechanism to realize the mixing process. Compared with other schemes, the process of mixed coin in Monero does not require the participation of users; any user can implement the mixed currency independently. Monero can effectively eliminate the denial-of-service attack on the decentralized coin scheme and assist with the problem of users' mixed coin leakage.

Mixed coin is widely used in the blockchain digital coin. There are many improvement schemes. We compare and analyze the schemes, as shown in Table \ref{tab0}.

\begin{table*}[!t]\scriptsize
\renewcommand{\arraystretch}{1.3}
\caption{Comparison of mix mechanisms in blockchain}
\label{tab0}
\centering
\begin{tabular}{|m{0.4in}<{\centering}|m{0.5in}<{\centering}|m{0.4in}<{\centering}|m{0.4in}<{\centering}|m{0.4in}<{\centering}|m{0.45in}<{\centering}|m{1.2in}|}
\hline
Reference&Protocol&Reliance on third parties&Risk of theft&Mixed coins cost&Resistance to DoS&Peculiarity \\
\hline
\cite{Chaum1981} & Mix & $\surd$ & $\surd$ & $\surd$ & Strong & The method is easy to use and is the most widely used. \\
\hline
\cite{BNM+14} & Mixcoin & $\surd$ & $\surd$ & $\surd$ & Strong & The proof can be raised to reduce the risk of theft. \\
\hline
\cite{VR15} & BlindCoin & $\surd$ & $\surd$ & $\surd$ & Strong & The blind signature mechanism is adopted to avoid leakage. \\
\hline
\cite{Dash18} & Dash & $\surd$ & $\surd$ & $\surd$ & Strong & The node that provides a mixed coin increases the cost of a violation by paying a deposit. \\
\hline
\cite{G13CoinJoin} & CoinJoin & N/A & N/A & N/A & Weak & No third parties involved, so there is no risk of theft. \\
\hline
\cite{RMK14} & CoinShuffle & N/A & N/A & N/A & Weak & The participants of the mixed coin do not know the details of the currency. \\
\hline
\cite{BOL+14} & Xim & N/A & N/A & $\surd$ & Strong & The method increases the difficulty of DoS attacks using a fee. \\
\hline
\cite{ZGH+15} & CoinParty & N/A & N/A & N/A & Strong & The mixing process can still function normally even if some participants violate the rules using the secure multi-party computation. \\
\hline
\cite{Monero18} & Monero & N/A & N/A & N/A & Strong & There is no need for multilateral negotiation using the ring signature mechanism. \\
\hline

\end{tabular}
\end{table*}

\subsubsection{\textit{Data Encryption}}

An encryption mechanism is a common scheme in the field of privacy protection. By encrypting sensitive data, users who hold secret keys can read the data, and others cannot decrypt it, even if they have access to it. Encryption ensures data privacy. In traditional blockchain, application data is stored in plaintext, and any node can access the data. Therefore, using encryption technology to protect privacy in blockchain must ensure that nodes can complete transaction verification tasks on encrypted data. In addition, since blockchain transactions must be jointly verified by all nodes, the impact of encryption mechanisms on validation efficiency must be reduced.

In blockchain, specific transaction information must be encrypted. In digital currency, there have been some protection schemes based on encryption.

\begin{itemize}

\item \textit{Monroe} \cite{Monero18} is an encrypted digital currency. In traditional digital currency, the content of the transaction output address is the receiver's public key and address information, and the observer can directly determine the coins' destination. In Monroe, the output address is the new address information obtained by the receiver's public key and the random parameter generated by the sender. Since the random parameter is only mastered by the sender, the observer cannot determine the relation between the new address information and the receiver. By generating different random parameters, it can be ensured that the output addresses of each transaction are different and there is no correlation between them. There are two key technologies in the Monroe coin, the stealth address and the ring signature. The stealth address is to address the problem of the relevance of the input/output address. A stealth address, while ensuring that the recipient's address changes every time, makes it impossible for an external attacker to see an address connection, but it does not guarantee anonymity between the sender and the receiver. Therefore, the Monroe coin developed a ring signature scheme. As shown in Figure \ref{fig10}, whenever a sender has to establish a transaction, he or she uses his or her private key and a certain number of public keys selected from other users' public key. to sign the transaction. When a signature is validated, the user must also use the other person's public key and parameters in his or her signature. At the same time, the sender must provide the key image to provide identity identification. Both the private and key images are once dense to ensure they cannot be traced.

    \begin{figure}[t]
    \centering
    \includegraphics[width=3in]{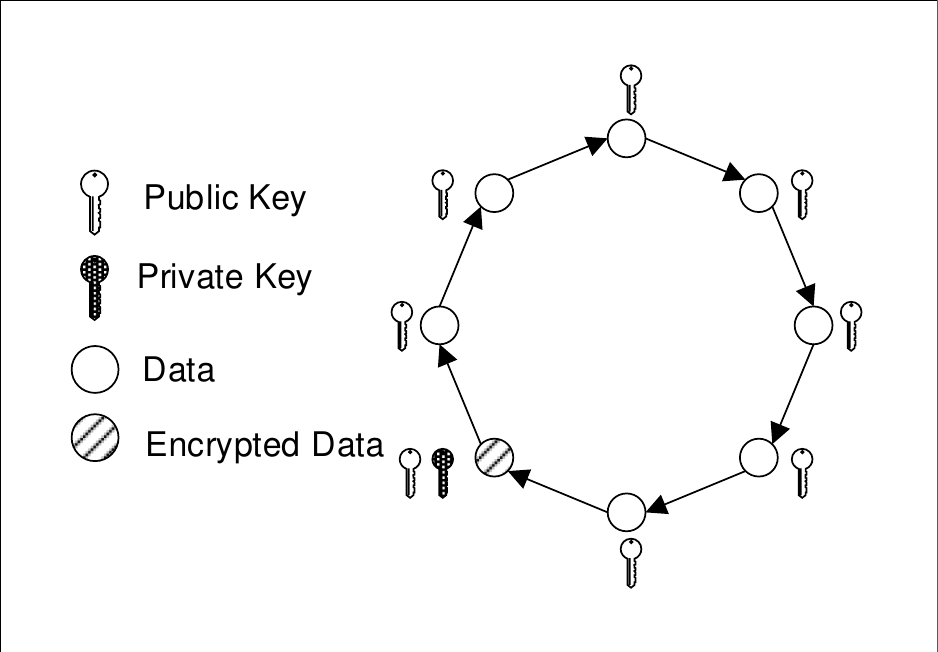}
    \caption{Ring signature diagram.}
    \label{fig10}
    \end{figure}

\item \textit{Zcash} \cite{SCG+14} is a new digital currency, formerly known as the Zerocoin \cite{MGG+13} project. It is an improvement on Zerocoin. Zcash uses the promise function to encapsulate the source of each transaction and the amount of several parameters, while using zk-SNARKs \cite{BCG+13} to prove the transaction. The proof process does not need to reveal relevant information; thus, it can hide the value of the sender and even the inputs and outputs of the transaction. Zcash is the best digital currency for privacy protection at present, but its adoption of the zk-SNARKs algorithm is very slow; it usually takes a minute to generate new proof and there is a bottleneck in efficiency. As shown in Figure \ref{fig11}, the underlying implementation is similar to the structure of Bitcoin, but Zcash is constructed using zk-SNARKs' decentralized mixed coin pool, and with the mint and pour operations it can perform in complete anonymity . Mint is the process by which a user writes a commitment to a list for a certain amount of cash. The promise must be a one-off serial number, and the user's private key is calculated and not reversible.

    \begin{figure}[t]
    \centering
    \includegraphics[width=3.5in]{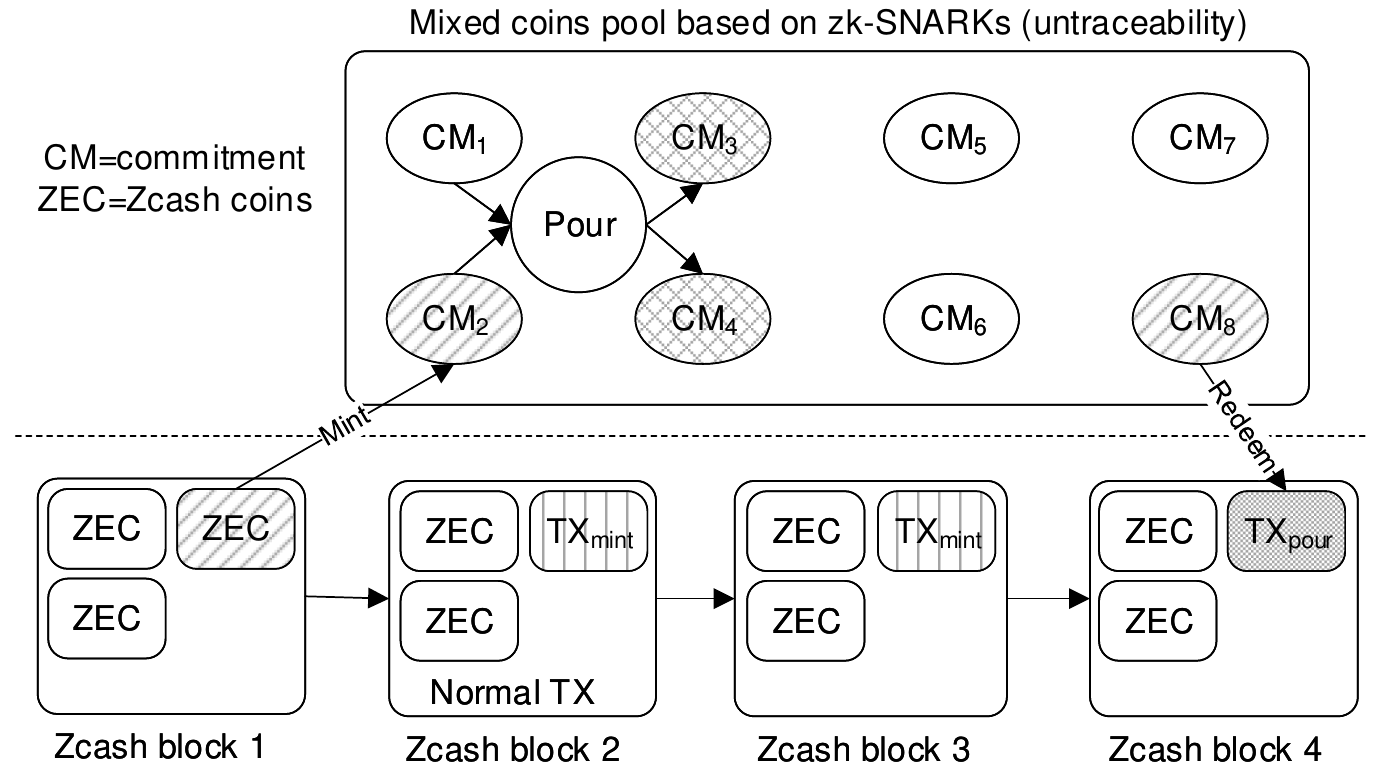}
    \caption{Zcash schematic diagram.}
    \label{fig11}
    \end{figure}

Similar to Bitcoin, the increase in the number of Zcash coin (ZEC) is based on mining. The ZEC obtained by a miner can be tracked and recorded, and its use also requires the signature of a private key. Therefore, if you use the ZEC directly, it is similar to Bitcoin, and you can directly complete the transfer between each address; however, it is currently not anonymous. The commitment made by the ZEC operation is not on the surface of the user address but depends on the public key and one one-time random number. When a user wants to spend (i.e., transfer) the ZEC, the user must provide the serial number and a commitment in the commitment list. In this way, the user can spend the ZEC without being completely exposed. The user can also use the redemption operation to extract the ZEC in the pool, the so-called redemption operation. The redemption is a commitment to return to a ZEC similar to the previous one, and the miner does not know which commitment was redeemed for the ZEC. Thus, you do not have to transfer ZEC to anyone, merely place a ZEC in the pool and redeem it; its source is untraceable.

\end{itemize}

\subsubsection{\textit{Restricted Release}}

The restriction release plan is to remove data that are directly related to privacy from the public database. Compared with the previous introduction of the mixed coin and encryption mechanisms, this type of method is completely guaranteed to ensure the security of privacy data. However, this approach has additional restrictions on business scenarios and requires additional modifications to the underlying protocol. Common solutions include the following:

\begin{itemize}

\item \textit{Lightning Network and Raiden Network:} The lightning network \cite{PD15} enables secure out-of-chain transactions. In the lightning network, the majority of the transaction details between users are implemented offline. Only the first and last transactions must be recorded on the blockchain ledger, so it can effectively protect transaction privacy. The Raiden network \cite{TH16} is a micropayment channel solution proposed by Ethernet. The Raiden network is directly based on the lightning network and has been developed. Because there are no specific field restrictions on the message format of the Ethernet smart contract, Raiden can introduce a single increment number for the channel balance snapshot, which solves the problem of identification and invalidation of the old version snapshot.
\item \textit{Consortium Blockchain and Private Blockchain:} Traditional blockchain applications are mostly public blockchain, such as BTC and ETH. In the public blockchain application, anyone is free to be a member of the blockchain network. The maintenance of the transaction data makes the public blockchain application highly credible, but it also brings the threat of identity and data privacy. To better protect privacy, blockchain technology produces a branch of consortium blockchain and private blockchain. Read and write permissions are open to nodes that join the alliance in the consortium blockchain. Read and write permissions are open to one node in the private blockchain.

\end{itemize}

\subsection{Data Availability Protection}

\subsubsection{\textit{Network Traceability Attacks Protection}}

The blockchain runs on a network with privacy protection so that its topology can be hidden, thereby preventing exposure of identity privacy information. Onion network (Tor) \cite{KPD+14,BAD+14} is one of the choices. Onion network is an anonymous communication technology, which protects the privacy of message sender and receiver and hides the route of data message passing through the network. Another is Monroe \cite{Monero18}. In traditional digital currency, the content of the transaction output address is the receiver's public key and address information, and the observer can directly determine the coins' destination. In Monroe, the output address is the new address information obtained by the receiver's public key and the random parameter generated by the sender.

\subsubsection{\textit{Eclipse Attacks Protection}}

Some researchers proposed several ways to solve the eclipse attack attack as follows.

\begin{itemize}

\item \textit{Restricting Access:} The network node needs to be authenticated. This method can effectively prevent the incoming and outgoing links of the node, so that the malicious node cannot access the blockchain node \cite{BAI15,AJA+15,EAA+15,DRN+14}. However,the approach will change the operational architecture of the blockchain.

\item \textit{Detecting and Blocking Malicious Nodes:} The blockchain uses a malicious node detection mechanism. Dillon et al. \cite{DJB14} proposesd an effective scheme for detecting malicious nodes and finds that malicious nodes add them to the blacklist, thereby limiting its further damage.

\end{itemize}

\subsection{Data Integrity Protection}

\subsubsection{\textit{Double-Spending Attack Protection}}

Karame et al. \cite{GOE12} analyzed that the current detection mode uses a ``listening period", which refers to the receiver detecting a collection of transactions after this period to determine whether double-spending exists. The problem with this approach is the attacker may delay the transmission of $TR_a$ because the neighbor node will not broadcast detected double-spending. Thus, $V$ cannot detect double-spending attacks even after the listening period. The fewer the number of neighbor nodes of the receiving node, the higher the success rate of this attack. Another method is to insert observers into the network, which immediately notify the receiver that double-spending has been detected. Only three observers can effectively detect double-spending attacks, but it requires additional costs. This paper proposed a mechanism to improve transactions' forwarding function, which is to forward to the neighbor node when a double-spending attack is detected. The detection rate of this mechanism is 100\%, with a false negative rate of 0\%.

Ruffing et al. \cite{TA15} designed a smart contract that allows payees to receive payments asynchronously and impose penalties on double-spending attackers. Eleftherios et al. proposed a new Byzantine consensus mechanism that shortend trading time by 15 to 20 seconds and used collective signatures to make transactions irreversible. George et al. \cite{GS15} proposed a new decentralized cryptocurrency RSCoin, the central bank maintains complete control over the coin supply to prevent double spending.

\subsubsection{\textit{Mining Attack Protection}}

Miners attack each other during mining to reduce the other party¡¯s or overall benefits. Yang et al. \cite{YMC17} proposed to establish a game model between two miners to improve the profit of miners through game strategies. When a loyal miner employs a pinning strategy, it can unilaterally set the payoff of a selfish miner within the range of zero to $r/2-c$ ($c$ is the computing power and $r$ is the expansion of profit), regardless of the selfish miner's strategy. The selfish miner's payoff is proportional to r but inversely proportional to $c$. The loyal miner cannot control his or her own payoff even with any subclass of the zero-determinant strategy.

Miller et al. \cite{AAJ+15} proposed the mining alliance mechanism, in which the members of the mining pool themselves did not trust each other, but submitted a password certificate to demonstrate the work they contributed. Shi \cite{S16} changed the consensus mechanism of Bitcoin, in which the value of N is established according to certain rules to ensure the continuous output rate of Bitcoin. The mechanism can improve dispersion and reduce the risk of 51\% attack. Gervais et al. \cite{AGV+16} analyzed the various parameters of PoW consensus mechanism. They designed the best countermeasures for double spending and selfish mining.

\subsection{Data Controllability Protection}

\subsubsection{\textit{Logic Problems of Smart Contracts}}

\begin{itemize}

\item \textit{Errors in Encoding State Machines:} The contract stipulates that in the absence of a winning player, the coins should be returned to their respective accounts. In Ethereum or Bitcoin, when multiple parties send input to a smart contract simultaneously, the order of transactions in this case depends on the miner mining the new block.
\item \textit{Failing to Use Cryptography:} Cryptography is the first line of defense for security protection. Application of cryptography is primarily to ensure binding and hiding. These two major attributes not only ensure that the transaction input is not tampered with but also ensure that the input information is not leaked. Before a message $M$ is committed, system need to compute a number that called $nonce$, $nonce$ and message $M$ hashed. The hash value is then sent as input to the contract.
\item \textit{Misaligned Incentives:} To ensure the correct behavior of all players, some incentives must be offered to ensure that players continue, regardless of winning or losing in accordance with the contract. In this scenario, a time limit can be established for the second player, and if the player submits the input before the time limit, another player will be rewarded. In addition, the game can also require both parties to pay a deposit in advance; if a player performs malicious behavior, the deposit is forfeited, bringing a loss to the player.
\item \textit{Ethereum-specific Mistakes:} For the Call-stack bug, developers are advised to minimize the use of variables so that functions are as small as possible. For the Blockhash bug, Luu et al. \cite{LDH16} solves the blockhash bugs allowing smart contract access to more than 256 blocks For Incentive bugs.

\end{itemize}

\subsubsection{\textit{Semantic Misunderstandings of Smart Contracts}}

Luu et al. \cite{LDH16} proposed to modify the operation semantics of Ethernet workshops to solve the above attack vulnerabilities, for example, guarded transactions (for TOD), deterministic timestamp and improved exception handling. However, in practical applications, all Ethernet clients must be upgraded. Then, a system called $OYENTE$ for detecting smart contracts is proposed, which can be used to verify the problems in an intelligent contract.

\subsubsection{\textit{Design Problems of Smart Contracts}}

Luu et al. \cite{LJR15} proposed a consensus-based computation protocol to solve the verifier's dilemma. They want to motivate miners to validate each transaction in each new block. If the miners did not verify the transaction in accordance with the stipulated protocol, they would receive few profits, and the honest miners who verified all the transactions would receive rewards that are unaffected by malicious miners.

\subsubsection{\textit{Privacy-Preserving Smart Contracts}}

\begin{itemize}

\item \textit{Privacy Preseration:} Blockchain is a completely open database; everyone can learn about each account asset and transaction record. Hawk \cite{KMS16} is a decentralized smart contract system, and to avoid the transaction plaintext details exposed in blockchain, Hawk programmers write smart private contracts without the implementation of cryptography. The compiler uses an encryption protocol (such as zero knowledge proof) to make the parties interact with the blockchain.

\item \textit{Data Feed:} Zhang et al. \cite{ZCC16} proposed a trusted link between non-blockchain applications and smart contracts,which called Town Crier (TC). It is a trusted link between smart contracts and non-blockchain applications. TC can help CSC deal with real world crimes (such as property crimes) \cite{AAE16}.

\end{itemize}

\section{Future Research Directions}\label{5}

The data security issues of blockchain are crucial to the future development of blockchain. Through the understanding and thinking of the research results of many scholars, we propose several future research directions.

\subsection{Data Privacy Protection Mechanism based on a Cryptology Algorithm}

An effective privacy protection scheme is to prevent an attacker from performing data analysis on the blockchain. However, this kind of scheme will change the underlying architecture of the blockchain, which is not conducive to use in the application. Therefore, it is necessary to design a scheme with high versatility. The solution should consider the computing and storage capabilities of the blockchain nodes.

\subsection{Data Availability Promotion Scheme based on Demand}

The existing anonymous attacks  are of low accuracy and high cost; they do not have the conditions for large-scale implementation, but the security threats of the data availability are universal in the blockchain, which uses P2P as the underlying protocol of the blockchain application, causing the hidden problems. In the future, we will focus on the appropriate access control policies to limit node access and malicious node detection mechanisms to prevent information leakage.

\subsection{Defenses Against PoW Attacks and New Consensus Mechanism}

Bitcoin's PoW consensus mechanism requires very strong computing power that have made mining by ``normal users" impractical. The collusion of miners or mines is very aggressive. An important research topic is methods to prevent the collusion of miners or mines.

In addition, proof of stake (PoS) \cite{KN12}, proof of personhood \cite{FJ08}, memory intensive \cite{ABF+14} and consensus alliance \cite{YGK+10} have been valued by many scholars.

\subsection{Verification of Smart Contracts}

Smart contracts often have problems with mismatch between expected and actual behavior. But the language of the verification tool limits its possibilities. Non-Turing complete and human-readable languages need further research, which is a future research direction. This is mentioned in \cite{AC14} \cite{FN16} \cite{BCG+16}.

\section{Summary}\label{6}

With the development of blockchain, its application is increasingly extensive, but the security of blockchain itself is gradually revealed. These problems pose a serious threat to blockchain and its application. In this survey, we created a comprehensive classification and summary of the security of blockchain. First, we present classification of blockchain attacks. Subsequently, we present the attacks and defenses of blockchain data in terms of the privacy, availability, integrity and controllability. Data privacy attacks present the threats of transaction and identity privacy. Data availability attacks present the threats brought by network traceability and eclipse attacks. Data integrity attacks present the threats brought by double-spending, selfish mining, and block withholding attacks. Data controllability attacks present the vulnerabilities of smart contracts. Finally, we provided several open research issues and provided some suggestions for the improvement of blockchain security.


\end{document}